\documentclass[journal]{IEEEtran}
\usepackage{eurosym}
\usepackage{color}
\usepackage{refcount}
\usepackage[numbers,sort&compress]{natbib}
\usepackage{setspace}
\usepackage{flushend}
\usepackage[caption = false]{subfig}
\usepackage{float}
\usepackage{fixltx2e}
\usepackage[normalem]{ulem} % either use this (simple) or

\usepackage[pdftex]{graphicx}
\DeclareGraphicsExtensions{.pdf, .pdf.bb, .eps}

\ifCLASSINFOpdf

\else
\fi
\usepackage[cmex10]{amsmath}
\def\mathclap#1{\text{\hbox to 0pt{\hss$\mathsurround=0pt#1$\hss}}}

\usepackage{amssymb}
\usepackage{varioref}
\usepackage{slashbox}

\usepackage{amsmath}% http://ctan.org/pkg/amsmath
\usepackage{algorithm,algcompatible}% http://ctan.org/pkg/{algorithms,algorithmicx}
\algnewcommand\algorithmicreturn{\textbf{return}}
\algnewcommand\RETURN{\algorithmicreturn}
\algnewcommand\algorithmicprocedure{\textbf{procedure}}
\algnewcommand\PROCEDURE{\item[\algorithmicprocedure]}%
\algnewcommand\algorithmicendprocedure{\textbf{end procedure}}
\algnewcommand\ENDPROCEDURE{\item[\algorithmicendprocedure]}%
\algnewcommand{\algvar}[1]{{\text{\ttfamily\detokenize{#1}}}}
\algnewcommand{\algarg}[1]{{\text{\ttfamily\itshape\detokenize{#1}}}}
\algnewcommand{\algproc}[1]{{\text{\ttfamily\detokenize{#1}}}}
\algnewcommand{\algassign}{\leftarrow}

\usepackage{makeidx}
\makeindex

\begin{document}

%\title{On the Security of Networked Control Systems in Smart Vehicles: Study of Covert   Attacks on Adaptive Cruise Control}
\title{A Note on the Security of ITS:  \\ Car Crash Analysis in Cruise Control Scenarios}
%Cyber Attack Prevention through  Zero False-Positive Rule-Set Generation for Industrial Systems Firewalls \\Prevention of Cyber Attacks on Industrial Systems through  Zero False-Positive   Rule-Set Generation for     Firewalls\\

\author{Mohammad~Sayad~Haghighi%,~\IEEEmembership{Senior Member,~IEEE}
\thanks{M. Sayad Haghighi is with 
the School of Science, Computing and Engineering Technologies, 
Swinburne University of Technology,  VIC 3122, Australia, Email: msayadhaghighi@swin.edu.au.}
}


\markboth{}%
{Shell \MakeLowercase{\textit{et al.}}: A Sample Article Using IEEEtran.cls for IEEE Journals}



\maketitle

\begin{abstract}
Security of Intelligent Transportation Systems (ITS) heavily depends on the security of the underlying components that create such a smart ecosystem. Adaptive Cruise Control (ACC) is embedded into most modern vehicles. In this report, we study the situations that the two vehicles involved in a cruise control scenario create. More precisely, after breaking down the phases the two vehicle go through (especially the ego one), we show how a simple formula can be used to predict collisions in hard brake cruise control scenarios.  
\end{abstract}


% \fi
\IEEEpeerreviewmaketitle

\section{Introduction}
The security of smart vehicles or intelligent transportation systems (ITS) in general, has become a source of concern for both consumers and industries \cite{ sardesai2018impacts,haghighiprotecting, farivar2019artificial,haghighi2019highly,sayad_persian2022}. ITS is gradually becoming part of the critical infrastructure. It has different components, ranging from in-vehicle modules to road side units \cite{abhishek2021drive} and inter-vehicle communication systems \cite{toorchi2013markov,fei2022digging,haghighi2010neighbor,haghighi2020intelligent}.  Malicious entities can compromise any of the above-mentioned units and jeopardize the safety of ITS \cite{mirzadeh2022filtering}. In this research, we explore the car crash issue when cruise control systems are targeted \cite{dantas2020formal,farivar2021security}. 
\section{Preliminaries\label{section:ACC}}
\subsection{ Adaptive Cruise Control}
We assume
the rear vehicle (ego) is equipped with an ACC module.  The front vehicle is  referred to as the lead. The goal of the ACC module is to make the ego  go at a speed set by the driver  as long as a safe distance is kept from the lead. This target speed is denoted by $v_{c}$. The ACC module {controls} the ego  in two  modes. In the first one, once $v_{c}$ is chosen by the driver, the  objective is to make the vehicle go at that speed as long as a safe distance is kept from the lead. The second mode is activated when the two vehicles get closer. In this mode, the ACC of the ego vehicle tries to keep it at a   safe distance from the lead. In short,
\begin {enumerate}
\item
$D_{rel}\geqslant D_{safe} \rightarrow$ (Track  $v_c$)

\item
$ D_{rel}<D_{safe} \rightarrow$   (Keep $D_{Safe}$) 
\end {enumerate}

Here, $D_{rel}$ is the relative distance of the ego and the lead vehicles and $D_{safe}$ is the safe distance.
The safe distance can be calculated as \cite{dekkata2019improved,farivar2021covert}:
\begin {align}
D=D_{default}+T_{gap}v_{e}
\label{Dsafe}
\end {align}
where $D_{default}$ is the default  spacing and $T_{gap}$ is a fixed value representing the necessary time gap.
\subsection{Stopping Distance Breakdown}
The distance a vehicle goes before it comes to a complete stop is determined by two factors; the reaction delay of the driver and the vehicle deceleration force.   During the first phase, the ego vehicle still goes by the same speed as the lead does (assuming that it has gone fast enough before to activate the second mode).    
\section{Crash Analysis\label{section:proposed_model}}

%\label{fig:stopping_dist}
%\end{left}
%\end{figure}

We  assume both vehicles are travelling at a speed of $v_c$ (in mode 2) and they are away from each other by a distance of $D$. If the lead suddenly brakes (e.g. as a result of seeing a barrier), at the time of brake, $v_l=v_e=v_c$ and $D(t=0)=D_{default}+T_{gap}v_{c}$.
The mathematical formulas for positions and velocities are as below. 
\begin{align}
v_{e}(t)=\begin{cases}v_c,~~~~~~~~~t\leq T_r \\
v_c+a_et ~~~ t>T_r \\
\end{cases}
\label{v_ego}
\end{align}
\begin{align}
x_{e}(t)=\begin{cases}x_e(0)+v_ct,~~~~~~~~~~~~~~~~~~~~t\leq T_r \\
x_e{(0)}+v_cT_r+\frac{1}{2}a_et^2+v_ct~~ t>T_r \\
\end{cases}
\label{x_ego}
\end{align}
%\begin{align}
%x_{e}(t)=\frac{1}{2} a_{e} t^2+v_c t+x_e(0)
%\end{align}  
in which $T_r$ is the ego driver's reaction time, $a_l$ and $a_e$ are the deceleration values which are negative and presumably vehicle-dependent. The air resistance impact on  $a_e$ and $a_l$ have been ignored by approximation. It is also assumed that the braking powers/decelerations remain constant until the vehicles stop.

Therefore, in the studied scenario, the lead and the ego are both moving at the speed of $v_c$, and are apart by a distance $D$. The time origin is considered to be the time the lead vehicle brakes. The path is considered straight and the location origin is assumed to be the front of the ego vehicle at the time of lead vehicle braking. The back of the lead vehicle participates in the equations (similar to the front of the ego).
The ego vehicle's driver responds to the front vechile's braking after $T_r$ seconds. 

\begin{figure}[t] %[b] %[p] puts at the end
   \centering
            \subfloat[]{  \includegraphics[width=\columnwidth]{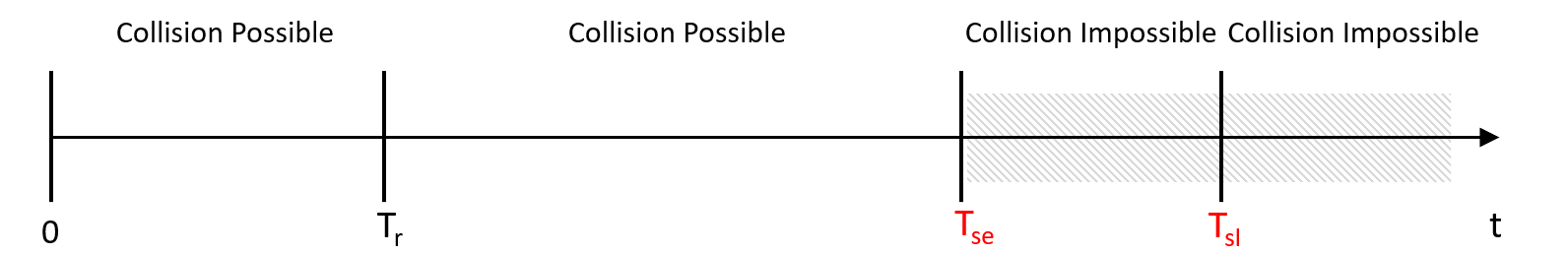}}
            \\
            \subfloat[]{  \includegraphics[width=\columnwidth]{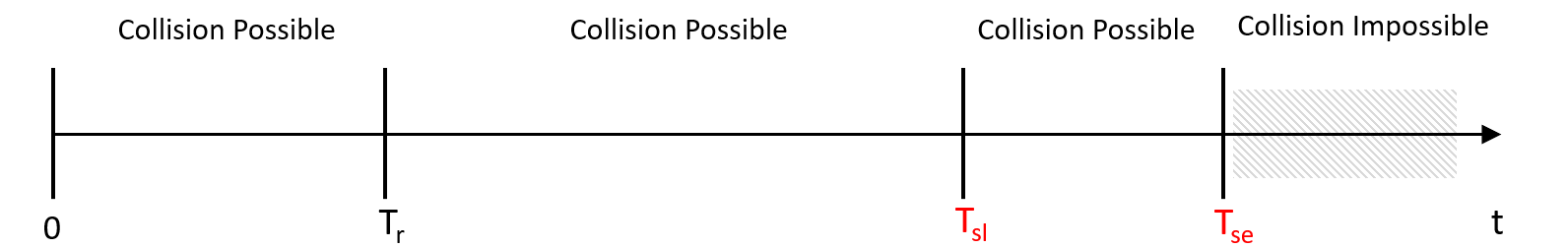}}
   \caption{Different phases two vehicles in a cruise control braking scenario go through (a) when the ego stops first at $T_{se}$ (b) when the lead stops first at $T_{sl}$. In this figure, $T_r$ stands for the ego driver's reaction time to the lead vehicle's braking. Note that $T_{sl}$ can theoretically lie in phase 1 too. But since the stop time of a vehicle is normally more than  a few seconds and is longer than an average driver's reaction time, this case has been ignored.}
\label{fig.brake}
\end{figure}

Fig.~\ref{fig.brake} shows the different phases the two vehicles go through. First, the lead vehicle brakes at $t=0$ (at a deceleration rate of $a_l$), but the ego vehicle does not change its speed since its driver needs $T_r$ seconds to respond. So in the first phase, the lead is decelerating and the ego is going at $v_c$. If the two do not collide in the first phase, then phase 2 begins, in which both ego and lead decelerate (at $a_e$ and $a_l$). Again in this phase, it is possible that the two collide, depending on the initial condition ($T_r$, $vc$ and $D$) and the deceleration values. If no collision occurs in this phase, another phase starts. But two cases can happen as phase 3. If the ego vehicle stops first (meaning that $T_{es}<T_{ls}$ as in Fig.~\ref{fig.brake}a), assuming that the collision has not happened yet, there is no possibility for it to happen ever. Since the lead is still moving thus going farther and further apart from the ego. So there is no chance for them to collide in phase 3. But if phase 3 starts with the lead stopping first (meaning that $T_{ls}<T_{es}$ as in Fig.~\ref{fig.brake}b), it is still possible to witness a collision. There is no phase 4 here, since after the ego vehicle stops, practically  no possibility of collision is left.

Let us study phase 1. A collision happens if the reaction time is too long so that ego catches up with the speed-decreasing lead. Mathematically speaking,

\begin{align}
v_ct=\frac{1}{2}a_lt^2+v_ct+D
\end{align}
which apparently yields the following results, 
\begin{align}
t_c&=\sqrt{-2D/a_l} ~~~ \text{(assuming $t_c \leq T_r$)} \label{eqphase1}\\
d_c&=v_{c}t_c
\end{align}

 However, we have a constraint here. The fact that we are seeking a potential collision in $t<T_r$ implies that there is a constraint on the deceleration of the lead vehicle. Mathematically, this translates to:

\begin{align}
&\frac{1}{2}a_lT_r^2+v_0T_r+D\leq v_0Tr\notag\\
&\frac{1}{2}a_lT_r^2+D \leq 0\notag\\
&\Rightarrow a_l\leq \frac{-2D}{T_r^2}\label{eq_phase1c}
\end{align}
But if $a_l>\frac{-2D}{T_r^2}$, collision will not happen in phase 1 rendering Eq.~\ref{eqphase1} invalid. 

Now, we shall study the possibility of collision in phase 2. Note that this phase is over if either of lead or ego stops.
For the time being, we assume the collision is going to happen during this phase. We have, 
\begin{align}
d_l(t)&=\frac{1}{2}a_lt^2+v_ct+D \label{eq.leadd}\\
d_e(t)&=\frac{1}{2}a_e(t-T_r)^2+v_c(t-T_r)+v_cT_r \label{eq.egoo}
\end{align}
at the collision time $d_l(t)=d_e(t)$. Therefore,
\begin{align}
&\frac{1}{2}a_lt^2+v_ct+D=\frac{1}{2}a_e(t-T_r)^2+v_c(t-T_r)+v_cT_r \notag\\
&2D+a_et^2=a_e(t-T_r)^2\notag\\
&(a_l-a_e)t^2+2a_eT_rt+2D-a_eT_r^2=0
\label{eq.time}
\end{align}
which has the following  roots if the condition $\Delta\ge0$ holds: 
\begin{align}
t_c&=\frac{-2a_eT_r\pm \sqrt{\Delta}}{2(a_l-a_e)} \label{eq.phase2} ~~~~\text{if $t_c\in[T_r,min(T_{sl},T_{se})]$}
\\\Delta&=4a_e^2T_r^2-4(a_l-a_e)(2D-a_eT_r^2)\notag\\
&=-4a_l (2D-a_eT_r^2)+8a_eD\\ \notag
&=8D(a_e-a_l)+4a_{l}a_{e}T_r^2\notag
\end{align}
The smallest positive answer  which is greater than $T_r$ determines the collision moment. If $\Delta <0$ or the responses fall outside the $[T_r,min(T_{sl},T_{se})]$ interval, the assumption that the two vehicles ever collide in phase 2 is invalidated. Here, $t_{sl}$ and $t_{se}$ denote the stopping times of the lead and the ego vehicle, respectively. These two can be found as below, 
\begin{align}
T_{sl}&=\frac{-v_c}{a_l} \label{eq.tstopl}\\
T_{se}&=\frac{-v_c}{a_e}+T_r\label{eq.tstop}
\end{align}

Of course the above hold if the there is no crash.  $T_{sl}$ can theoretically lie in phase 1 too. But since the stop time of a vehicle is normally more than  a few seconds (at relatively high cruise speeds) and is longer than an average driver's reaction time, we can assume the stopping happens in phase 2.

% \begin{figure*}[h] %[b] %[p] puts at the end
% %\begin{left}
% %   \centering
%    \subfloat {\includegraphics[width=1\columnwidth, height=0.9\columnwidth]{covert1_IDS_withoutComp.pdf}}
%    \subfloat{\includegraphics[width=1\columnwidth, height=0.9\columnwidth]{covert2_IDS_withComp.pdf}}
%     %   \vspace{-2ex}
% \caption{The intelligent intrusion detection system for the first scenario which the attack time is occurred at  $40$ sec and the detection is the same as the attack time.  (Left) There is not any compensation strategy. (Right) The P Controller is applied to compensate the malicious attack occurred into the ACC system. }
% \label{fig.result_scen2}
% %\end{left}
% \end{figure*}
% 
% 
% 

If either of the responses to Eq. \ref{eq.phase2} falls within $[Tr,min(T_{sl},T_{se})]$, the collision happens in phase 2. In the unlikely event that both fall within the interval, the smaller value should be considered as the collision time.  But if none of the responses/roots fall into this interval,  we have to search for possible collisions in phase 3.

Now that the collision has not happened in phase 1 and 2, we have to check the possibility of collision in phase 3. However, phase 3 is somewhat irregular, meaning that the start time of it depends on the relative stopping times of the two vehicles. 
If $T_{se}<T_{sl}$, assuming that collision has not happened in the previous two phases, no collision will happen anymore. Because once the ego stops, the lead moves further away and the chance of collision becomes zero in this case. 
However, if $T_{sl}<T_{se}$, there is still a window of possibility for collision. To find the time of collision in this phase, we should intersect the still position of the leading vehicle with the location equation of the moving ego. Given Eq. \ref{eq.tstopl} and from the time equation of the lead's location we have, 
\begin{align}
d_{sl}=\frac{-v_c^2}{2a_l}+D
\end{align}

by intersecting Eq. \ref{eq.egoo} with the above we will have,

\begin{align}
\frac{-v_c^2}{2a_l}+D=\frac{1}{2}a_e(t-T_r)^2+v_c(t-T_r)+v_cT_r\notag\\
\Rightarrow a_et^2+2(v_c-a_eT_r)t+a_eT_r^2+\frac{v_c^2}{a_e}-2D=0 \label{eq.phase33}
\end{align}
which can be solved easily by following the same approach as we did for Eq. \ref{eq.time}. $t_c$ will be the answer (out of the two) which is greater than $T_{sl}$. In case both are, the minimum of the two will be the answer. However, if  If there is no real solution for Eq. \ref{eq.phase33},  then the vehicles do not collide at all. This is the final phase.

For phase 3, if we are merely interested in knowing whether the two vehicles have crashed or not, we can use simpler equations. Apparently, being in phase 3 implies that the vehicles have not crashed in phase 1 and 2. Therefore, the lead is still ahead (despite being still) and the ego vehicle is behind and moving. To check if the two will crash, it is easier to check if the stopping distance of the ego vehicle eventually becomes greater than that of the lead. Therefore, the two vehicles collide in phase 3 if the following condition is satisfied. 
\begin{align}
v_cT_r-\frac{v_c^2}{2a_e}>   -\frac{v_c^2}{2a_l}+D \label{phase3.collide}
\end{align}
This is a useful equation, however, this does not give the time or location of the probable accident.

For the two cars not to crash, none of the conditions for collision in the three phases should be met. This means that Eq. \ref{eq_phase1c} must not hold so as to a collision in phase 1 is avoided. Moreover, Eq. \ref{eq.time} must not have real answers, and if it does, they must fall outside the $[T_r,min(T_{sl},T_{se})]$ interval. Finally, if the lead stops first (after calculating Eq. \ref{eq.tstopl} and Eq.~\ref{eq.tstop}), Eq.~\ref{phase3.collide} must not hold. If all of these conditions are met, the two vehicles do not collide.

As seen, this is a rather complex and non-straight forward process. We already know that Eq.~\ref{phase3.collide} is determinant in phase~3. Given that cruising speeds are normally high and the safety distance (D) is reasonably large (calculated below), the probability of having a collision in the first phase is almost zero for a normal driver's reaction time.  
\begin {align}
D_{}=D_{default}+T_{gap}v_{e}
\label{Dsafe}
\end {align}
This formula suits dry roads with $D_{default}=10m$, $T_{gap}=1.4s.$. Apparently,  in a cruise control scenario, $v_e=v_c$.

To better understand why phase 1 collisions are very rare, we use a dataset provided by performancedrive.com.au
 \cite{datasetc}. This Australian source reviews vehicles and measures a few performance indicators like the stopping distance (at 100km/h) from which the maximum deceleration can be estimated. The dataset includes around 600 vehicles which are  driven  in Australia.
We refer to this dataset as PD henceforth.

At 100km/h, $D=48.9m$, and the range of deceleration for a variety of car makes in PD is $[-11.91, -7.7] m/s^2$.
For a collision not to happen in phase 1, $a_l\leq \frac{-2D}{T_r^2}$ must not hold. This means that $T_r<\sqrt{\frac{-2D}{a_l}}$.  For the above range of potential $a_l$ values, this condition translates to $T_r<2.9s$ in the worst case scenario (i.e. $a_l=-11.91$). This value is larger for other normal cars in the dataset. This reaction time is much longer than the average, thus making collision in the first phase improbable. This deduction does not change much even for somewhat lower speeds, considering that normal reaction times are around or below 1s. 

Collisions in phase 2 are more complicated. Eq. \ref{eq.phase2}, if holds, gives the collision time. Since we have established that collisions will (almost certainly) not happen in phase 1, we may use Eq. \ref{eq.leadd} and Eq. \ref{eq.egoo} till the vehicles stopping times. We argue that a collision can (approximately) be predicted in this phase by checking if the hypothetical stopping distance of the ego is greater than that of the lead. This gives us the same relation as Eq. \ref{phase3.collide}. However, this time we are using it for both phase 2 and phase 3. In phase 3, this relation makes error-free predictions. 

\begin{figure}[h] %[b] %[p] puts at the end
%\begin{left}
   \centering
   \vspace{-1.5ex}
            \subfloat[]{  \includegraphics[width=0.9\columnwidth]{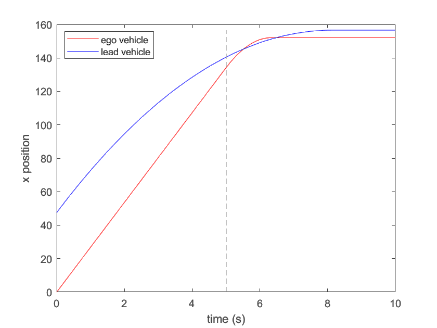}}\\
                 \subfloat[]{     \includegraphics[width=.9\columnwidth]{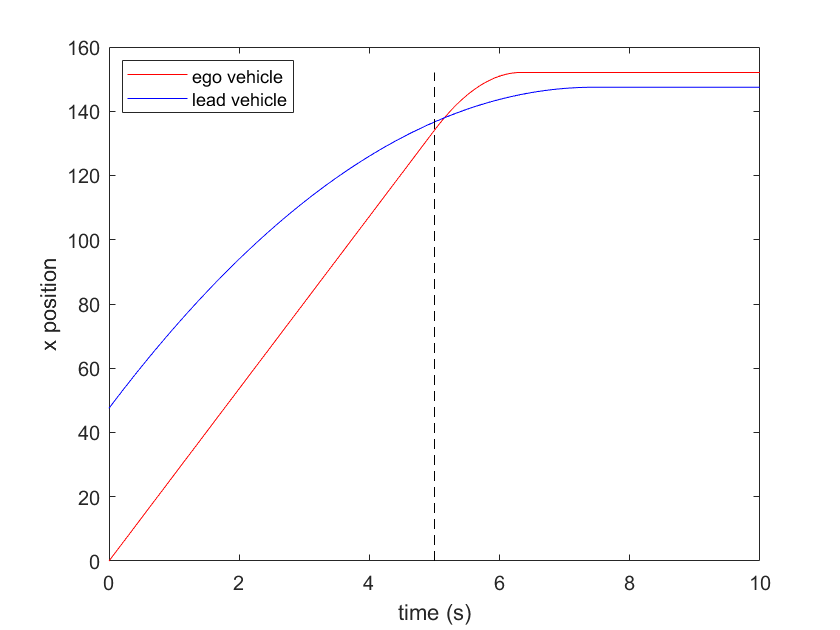}}\\
                 \subfloat[]{     \includegraphics[width=.9\columnwidth]{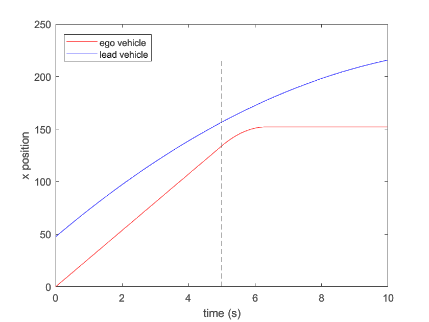}}\\
   \caption{Demonstration of the lead and ego vehicle positions over the time for different deceleration values: (a) $a_l=-3.3$, $a_e=-20$, $T_r=5$, (b) $a_l=-4$, $a_e=-20$, $T_r=5$, (c) $a_l=-2$, $a_e=-20$, $T_r=5$. In the above, $v_c=96.6$, $T_{gap}=1.4$, and  $D_{default}=10$.}
\label{fig:response_area2}
%\end{left}
\end{figure}

However, in phase 2, despite not being obvious, there are exceptions.
Exception happens when the reaction time is too long and the deceleration power of the ego vehicle is significantly larger than that of the lead. For example, the lead vehicle brakes at $t=0$ and its speed decreases linearly with time afterwards. Its position changes with $t^2$ though. In the meantime, the ego vehicle is still following the previous speed, since the reaction time is long. Apparently, the two vehicles get closer, because previously they were traveling at the same speed and they were only apart by $D$. Now that the lead is braking, their distance gets closer than $D$. This can continue, even in the initial moments that the ego vehicle also brakes, till the ego vehicle's position becomes equal to  (and in mathematics, greater than) that of the lead. In an imaginary scenario where these two cars run in parallel lines, after this incident, the ego vehicle brakes, and since its deceleration rate is very high, it will stop soon, but the lead's deceleration value is small thus it will take longer for it to stop, making its final stopping position somewhat greater than that of the ego vehicle. In such a complicated case,  case, Eq. \ref{phase3.collide} misses the collision, but as explained this phenomenon can indeed happen. The question is how probable is this?

We tested this approximate formula for the prediction of collision (i.e. Eq. \ref{phase3.collide}) on the PD dataset. This dataset includes a variety of vehicles running in the roads and streets of Australia. 

We swept all the car makes in the 600-vehicle dataset, and tested each with every other vehicle as (lead, ego) pairs. The hypothesis being tested was whether Eq. \ref{phase3.collide} yields the same collision predictions as the precise 3-phase  approach or not.

The above two approaches were run with
$v_c = 27.78m/s$ 
$T_r=\{0,0.5,1,1.5,2,2.5\}s$, $Tgap=1.4s$, and $D_{default}=10m$. In the almost 360,000 experiments conducted, zero inconsistency was reported. Meaning that the simplified model of Eq.~\ref{phase3.collide} can predict collisions in practical settings as efficiently as the precise multi-phase model does. 

But  this does not mean the exceptions we mentioned cannot happen. We deliberately changed the settings manually to at least find one sample. Fig.~\ref{fig:response_area2} demonstrates the lead and ego vehicle positions over the time for different deceleration values. In Fig. \ref{fig:response_area2}(a) the phenomenon has happened in phase~2. The stopping position of the ego is shorter than that of the lead but collision has happened. But a careful review of the settings shows why this  never happened with the real-world settings of the PD dataset. We could produce such an exception with $T_r=5$,  $a_l=-3.3$ and  $a_e=-20$ at a speed of  $v_c=96.6km/h$ (with  $T_{gap}=1.4$ and  $D_{default}=10$).
Apparently, the range of decelerations is very extreme, and so is the reaction time. Even a slight change of the deceleration values will not produce such exceptions. This confirms our guess about the rarity of the exceptions in real-world scenarios.

%\section{Conclusion\label{section:conclusion}}
%We 

\bibliographystyle{IEEEtran}
%{\small \bibliography{referencesR1}}

%\bibliographystyle{plain}
\bibliography{references}

\end{document}